\documentclass[prd,twocolumn,showpacs,floatfix,amsmath,nofootinbib,amssymb,floatfix]{revtex4}

\usepackage{graphicx,color,dcolumn,booktabs,bm,multirow}
\usepackage{longtable,lscape}
\usepackage{txfonts}
\usepackage{overpic}
\usepackage{amssymb}
\usepackage{indentfirst}
\usepackage{feynmf}   
\usepackage{slashed}  
\usepackage{cases}
\usepackage{color}
\usepackage{multirow}
\usepackage{epstopdf}

\usepackage{ulem} 

\def\lrpartial{\buildrel\leftrightarrow\over\partial}

\begin{document}
\title{Radiative and pionic transitions from the $D_{s1}(2460)$ to the $D_{s0}^\ast(2317)$}
\author{Cheng-Jian Xiao$^{1,4}$}
\author{Dian-Yong Chen$^{1,2}$\footnote{Corresponding author}}\email{chendy@impcas.ac.cn}
\author{Yong-Liang Ma$^{3}$}
\affiliation{$^1$Institute of Modern Physics, Chinese Academy of Sciences, Lanzhou 730000, China\\
$^2$Research Center for Hadron and CSR Physics, Lanzhou
University $\&$ Institute of Modern Physics of CAS, Lanzhou 730000, China\\
$^3$ Center of Theoretical Physics and College of Physics, Jilin University, Changchun, 130012, China\\
$^4$ University of Chinese Academy of Sciences, Beijing 100049, China}

\date{\today}
\begin{abstract}
We estimate the partial widths for the radiative and pionic transitions from the $D_{s1}(2460)$ to the $D_{s0}(2317)$ in a molecular scenario, in which the $D_{s1}(2460)$ and $D_{s0}^\ast(2317)$ are considered as hadronic molecular states of $DK$ and $D^\ast K$, respectively. The partial widths for the $D_{s1}(2460) \to D_{s0}^\ast(2317) \pi^0$ and $D_{s1}(2460) \to D_{s0}^\ast(2317) \gamma$ are evaluated to be about $0.19\text{--}0.22$ and $3.0\text{--} 3.1$ keV, respectively. In addition, the ratio of the $D_{s1}(2460) \to D_{s0}(2317) \gamma$ and $D_{s1}(2460) \to D_{s}^\ast \pi^0$ is estimated to be about $(6.6\text{--}10.6) \times 10^{-2}$, which is safely under the measured upper limit.
\end{abstract}
\pacs{14.40.Pq, 13.20.Gd, 12.39.Fe}

\maketitle

\section{Introduction}\label{sec1}

In the last decade, great experimental progress in the charmed-strange meson spectrum has been achieved. Some radially or orbitally excited charmed-strange mesons have been observed \cite{Agashe:2014kda}, and these observations not only make the charmed-strange meson family lengthy, but also raise some challenges to the conventional quark model  \cite{Godfrey:1985xj}. Among these newly observed charmed-strange mesons, the $D_{s0}^{\ast}(2317)$ and $D_{s1}(2460)$ are two particular states, since their masses are far below the quark model expectations \cite{Godfrey:1985xj}.

The $D_{s0}^\ast(2317)$ was first reported by the {\it BABAR} Collaboration in the $D_s^+ \pi^0$ invariant mass spectrum of the $B$ decay process and its mass was measured to be $(2316.8 \pm 0.4)$ MeV \cite{Aubert:2003fg}. Later, the CLEO Collaboration confirmed the existence of this state and also reported another state, the $D_{s1}(2460)$, in the $D_s^{\ast +} \pi^0 $ invariant mass distribution, which is $351.2 \pm 1.7 (\mathrm{stat}.) \pm 1.0 (\mathrm{syst}.)$ MeV heavier than the $D_{s}^\ast$ \cite{Besson:2003cp}. Besides the $D_s^\ast \pi^0$ mode, some other decay modes of the $D_{s1}(2460)$--like $D_s \gamma$, $D_s^{\ast} \gamma$, $D_{s} \pi^+ \pi^-$ and $D_{s0}^\ast \gamma $--have also been measured~\cite{Besson:2003cp}.

After the observations from the {\it BABAR} and CLEO collaborations \cite{Aubert:2003fg, Besson:2003cp}, the existence of the $D_{s0}^\ast(2317)$ and $D_{s1}(2460)$ was confirmed by the Belle Collaboration \cite{Abe:2003jk, Krokovny:2003zq} and {\it BABAR} Collaboration\cite{Aubert:2004pw, Aubert:2003pe, Aubert:2006bk}. And now, the Particle Data Group (PDG) averages of the masses of the $D_{s0}^\ast(2317)$ and $D_{s1}(2460)$ are \cite{Agashe:2014kda}
\begin{eqnarray}
m_{D_{s0}^\ast}(2317)&=& (2317.7 \pm 0.6)\ \mathrm{MeV}, \nonumber\\
m_{D_{s1}}(2460)&=&(2459.5 \pm 0.6)\  \mathrm{MeV}.\nonumber
\end{eqnarray}
In addition, the decay mode of $D_{s1}(2460) \to D_{s0}^\ast(2317) \gamma$ was measured by the CLEO and {\it BABAR} collaborations \cite{Besson:2003cp, Aubert:2003pe} and  the ratio of $\Gamma(D_{s1}(2460) \to  D_{s0}^\ast(2317) \gamma ) $ and $ \Gamma(D_{s1}(2460) \to D^\ast \pi^0)$ was reported to be
\begin{eqnarray}
\frac{\Gamma(D_{s1}(2460) \to  D_{s0}^\ast(2317) \gamma )}{ \Gamma(D_{s1}(2460) \to D_s^\ast \pi^0)}
\left\{
  \begin{array}{ll}
    <0.58, & \hbox{CLEO \cite{Besson:2003cp},} \\
    <0.22, & \hbox{{\it BABAR} \cite{Aubert:2003pe},}
  \end{array}
\right.
\label{Eq:ratio}
\end{eqnarray}
at the $90 \%$ confidence level.

Theoretically, the quark model predicted the masses of the $D_{s0}^\ast$ and $D_{s1}$ to be 2480 MeV and $2530$~MeV \cite{Godfrey:1985xj}, respectively, which are about $160$ and $70$~MeV, respectively, above the experimental measured values. This disagreement between the quark model expectations and experimental measurements makes these two states unlike conventional charmed-strange mesons.

The particular properties of the $D_{s0}^\ast(2317)$ and $D_{s1}(2460)$ have stimulated the theorists' interest in the nature of these two states. The coupled channel estimates indicated that the masses of the $D_{s0}^\ast (2317)$ and $D_{s1}(2460)$ could result from the strong coupling of the $P\text{-}$wave charmed-strange mesons to the $DK$ and $D^\ast K$, respectively \cite{Lutz:2008zz, Hwang:2004cd}. With some fine-tuning parameters, the masses of the $D_{s0}^\ast(2317)$ and $D_{s1}(2460)$ could be reproduced in a relativistic quark model \cite{Liu:2013maa}. The decays of the $D_{s0}^\ast(2317)$ and $D_{s1}(2460)$ were investigated in a conventional charmed-strange mesons frame with different methods, such as the quark pair-creation model \cite{Liu:2006jx, Lu:2006ry}, QCD sum rules \cite{Wang:2006mf, Dai:2003yg, Colangelo:2003vg, Colangelo:2005hv, Colangelo:2012xi}, and chiral effective theory \cite{Fajfer:2015zma}. However,
the large-$N_c$ expansion calculations indicated that the $D_{s0}^\ast(2317)$ could not be a standard quark-antiquark meson~\cite{Guo:2015dha}. A $c\bar{s}q\bar{q}$ tetraquark interpretation was proposed to understand the mass and decay behavior of the $D_{s0}^\ast(2317)$ \cite{Nielsen:2007zz, Nielsen:2005zr, Terasaki:2003qa}. The QCD sum rule calculations also supported the idea that the $D_{s0}^\ast(2317)$ could be a tetraquark state \cite{Wang:2006uba, Bracco:2005kt}.

Since the masses of the $D_{s0}^\ast(2317)$ and $D_{s1}(2460)$ are about 40 MeV below the thresholds of the $D K$ and $D^\ast K$, respectively, a possible explanation of the structures of $D_{s0}^\ast(2317)$ and $D_{s1}(2460)$ is that they are $DK$ and $D^\ast K$ hadronic molecules, respectively. The calculations in the Bethe-Salpeter approach \cite{Xie:2010zza} and potential model \cite{Zhang:2006ix} showed that the $D_{s0}^\ast(2317)$ could indeed be a $DK$ hadronic molecule. In Ref. \cite{Bicudo:2004dx}, the $D_{s0}^\ast(2317)$ and $D_{s1}(2460)$ were considered as kaonic molecules bound by strong short-range attraction. The decay behaviors of the $D_{s0}^\ast(2317)$ and $D_{s1}(2460)$ were extensively investigated in the $DK$ and $D^\ast K $ hadronic molecular scenario \cite{Faessler:2007gv, Faessler:2007us, Cleven:2014oka}. The production of the $D_{s0}^\ast(2317)$ and $D_{s1}(2460)$ from the nonleptonic $B$ decay were calculated in Ref.~\cite{Datta:2003re}, in which $D_{s0}^\ast(2317)$ and $D_{s1}(2460)$ were considered as hadronic molecular states of $DK$ and $D^\ast K$, respectively .

In this paper, we study the radiative and pionic transitions from the $D_{s1}(2460)$ to the $D_{s0}^\ast(2317)$ in a hadronic molecular scenario. With the assignment that the $D_{s0}^\ast(2317)$ and $D_{s1}(2460)$ are the hadronic molecules of $DK$ and $D^\ast K$, respectively, one could find that the radiative and pionic transitions from the $D_{s1}(2460)$ to the $D_{s0}^\ast (2317)$ occur via the subprocesses $D^\ast \to D\gamma$ and $D^\ast \to D \pi^0$, respectively. As for the $D_{s1}(2460) \to D_{s0}^\ast(2317) \pi^0$, it is an isospin-violating process, which could result from the mass differences of charged and neutral $D$ and $K$ mesons and $\eta-\pi^0$ mixing. In addition, the ratio of the partial widths for the $D_{s1}(2460) \to D_{s0}^\ast(2317) \gamma$ and $D_{s1}(2460) \to D^\ast  \pi^0$ was measured by the CLEO and {\it BABAR} collaborations \cite{Besson:2003cp, Aubert:2003pe}. In the present work, we can test the $D^\ast K$ assignment of the $D_{s1}(2460)$ by comparing the estimated ratio of $\Gamma(D_{s1}(2460) \to  D_{s0}(2317) \gamma )$ and $\Gamma(D_{s1}(2460) \to D^\ast \pi^0)$ with the experimental measurements.

This work is organized as follows. The the hadronic molecular structures of the $D_{s0}^\ast(2317)$ and $D_{s1}(2460)$ are discussed in Sec.~\ref{Sec2}. The partial widths for $D_{s1}(2460) \to D_{s0}^\ast(2317) \pi^0,\ \ D_{s0}^\ast(2317) \gamma \ {\text{and}}\ D_s^\ast \pi^0$ are estimated in Sec.~\ref{Sec3}. The numerical results are presented in Sec.~\ref{sec4} and Sec.~\ref{sec5} is dedicated to a short summary.

\section{Hadronic molecular structures of the $D_{s0}^\ast(2317)$ and $D_{s1}(2460)$}\label{Sec2}

In the hadronic molecular scenario, the $D_{s0}^\ast(2317)$ and $D_{s1}(2460)$ are assigned as $S$-wave $DK$ and $D^\ast K$ hadronic molecules, respectively. Here, we adopt the following effective Lagrangians to describe the interactions of the $D_{s0}^\ast(2317)$ and $D_{s1}(2460)$ and their constituents. The concrete forms of the Lagrangians are \cite{Faessler:2007gv,Faessler:2007us}
\begin{eqnarray}
\label{eq:s0dk}
\mathcal{L}_{D_{s0}^*}(x)&=&g_{D_{s0}^\ast DK}\,D_{s0}^{\ast }(x)\int dy\,\Phi_{D_{s0}^\ast }(y^2)\,D^{T}(x+w_{KD}y)\nonumber\\
&&\times \,K(x-w_{DK}y)+\text{H.c.}\, ,\\
\label{eq:s1dstark}
\mathcal{L}_{D_{s1}}(x)&=&g_{D_{s1}D^\ast K}\,D_{s1}^{\mu}(x)\int dy\,\Phi_{D_{s1}}(y^2)\,D^{\ast T}_{\mu}(x+w_{KD^\ast }y)\nonumber\\
&&\times \,K(x-w_{D^\ast K}y)+\text{H.c.}\, ,
\end{eqnarray}
where
$$D^{(\ast )T}=(D^{(\ast) 0},D^{(\ast) +}),\quad K=\left(\begin{array}{c} K^{+} \\ K^{0}\end{array}\right).$$
The $w_{ij}=m_i/(m_i+m_j)$ is kinematical parameter with $m_{i}$ being the mass of the corresponding meson.

The correlation functions $\Phi_{{D_{s0}^\ast }}(y^2)$ and $\Phi_{{D_{s1}}}(y^2)$, which depend only on the Jacobian  coordinate $y$, are introduced to depict the distributions of the components in the hadronic molecule. The Fourier transformation of the correlation function is,
\begin{equation}
\Phi_M(y^2)=\int \frac{d^4p}{(2\pi)^4}\, e^{-ipy}\,\tilde\Phi_M(-p^2,\Lambda^2_M),\quad M=(D_{s0}^\ast ,\,D_{s1}).
\end{equation}

The introduced correlation function also plays the makes the
Feynman diagrams finite in the ultraviolet region of Euclidean space, which indicates that the Fourier transformation of the correlation function should drop fast enough in the ultraviolet region. Here we choose the Fourier transformation of the correlation in the Gaussian form,
\begin{equation}
\tilde\Phi_{M}(-p^2,\Lambda_M^2)=\text{exp}\,(p^2/\Lambda^2_{M}),\quad M=(D_{s0}^\ast ,\,D_{s1}),
\end{equation}
with $\Lambda_M$ being the size parameter which characterizes the distribution of components inside the molecule.

\begin{figure}[htb]
\begin{tabular}{cc}
\includegraphics[scale=0.5]{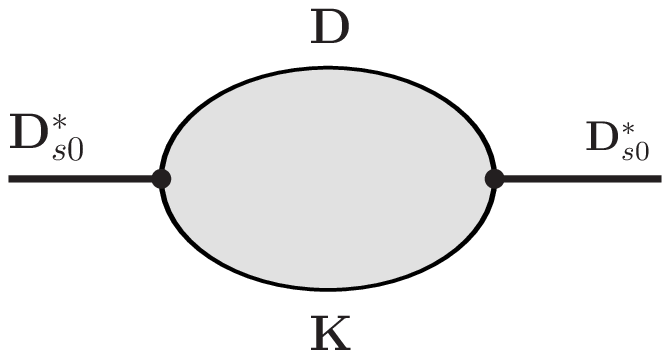}&
\includegraphics[scale=0.5]{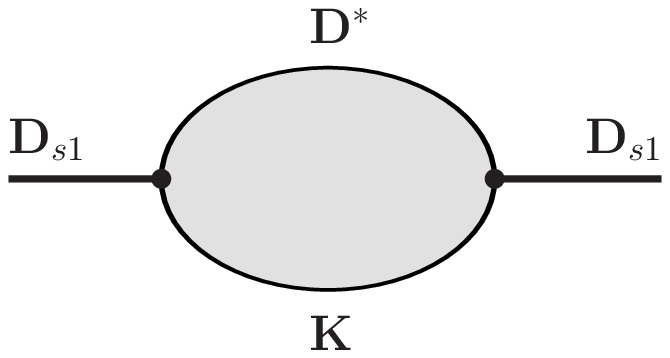}\\
$(a)$  & $(b)$
\end{tabular}
\caption{Mass operators of the $D_{s0}^\ast(2317)$ (a) and $D_{s1}(2460)$ (b). \label{fig:m-o}}
\end{figure}

The coupling constants $g_{D_{s0}^\ast DK}$ and $g_{D_{s1}D^\ast K}$ in Eqs.~(\ref{eq:s0dk}) and (\ref{eq:s1dstark}) could be determined by
the compositeness conditions \cite{Weinberg:1962hj, Salam:1962ap, Hayashi:1967,Faessler:2007gv,Faessler:2007us}, where the renormalization constants of the composite particles should be zero, i.e.,
\begin{eqnarray}
\label{eq:compsitness-con}
Z_{D_{s0}^\ast }&\equiv&1-\Sigma^\prime _{D_{s0}^\ast }(m^2_{D_{s0}^\ast })=0,\nonumber\\
Z_{D_{s1}}&\equiv&1-\Sigma^\prime _{D_{s1}}(m^2_{D_{s1}})=0,\label{eq:cp1}
\end{eqnarray}
with $\Sigma^\prime _{D_{s0}^\ast }(m^2_{D_{s0}^\ast })$ being the derivative of the mass operator of the $D_{s0}^\ast(2317)$. As for the $D_{s1}(2460)$, the mass operator $\Sigma^{\mu\nu}_{D_{s1}}$ presented in Fig. \ref{fig:m-o}(b) can be decomposed into the transverse $\Sigma_{D_{s1}}$ and longitudinal $\Sigma_{D_{s1}}^L$ components as \begin{equation}
\label{eq:xs1-compsitness-con}
\Sigma^{\mu\nu}_{D_{s1}}(p)=g^{\mu\nu}_{\perp}
\Sigma_{D_{s1}}(p^2)+\frac{p^{\mu}p^{\nu}}{p^2}
\Sigma_{D_{s1}}^{L}(p^2),
\end{equation}
with $g^{\mu\nu}_{\perp}=g^{\mu\nu}-p^{\mu}p^{\nu}/p^2$. The concrete forms of the mass operators of the $D_{s0}^\ast(2317)$ and $D_{s1}(2460)$ corresponding to the diagrams in Fig.~\ref{fig:m-o} are
\begin{eqnarray}
\Sigma_{D_{s0}^\ast} &=& g_{D_{s0}^\ast DK}^2\int \frac{d^4q}{(2\pi)^4}
    \tilde\Phi^2[-(q-w_{DK}p)^2,\Lambda^2]\nonumber\\
&&\times\frac{1}{(p-q)^2-m_K^2}\frac{1}{q^2-m_D^2},\\
\Sigma_{D_{s1}}^{\mu \nu} &=& g_{D_{s1} D^\ast K}^2\int \frac{d^4q}{(2\pi)^4}
    \tilde\Phi^2[-(q-w_{D^\ast K} p)^2,\Lambda^2]\nonumber\\
&&\times\frac{1}{(p-q)^2-m_K^2}\frac{-g^{\mu\nu}+{q^{\mu}q^{\nu}/m^2_{D^{\ast}}}}{q^2-{m^2_{D^{\ast}}}}.
\end{eqnarray}

\section{Radiative and pionic transitions from the $D_{s1}(2460)$ to the $D_{s0}^\ast (2317)$}\label{Sec3}

We estimate the partial widths for the radiative and pionic transitions from the $D_{s1}(2460)$ to the $D_{s0}^\ast (2317)$ in an effective Lagrangian approach. The interactions of the $D_{s0}^\ast (2317)$ and $D_{s1}(2460)$ with their components are presented in Eqs. (\ref{eq:s0dk}) and (\ref{eq:s1dstark}). Besides these effective Lagrangians, in our calculation, we employ the following phenomenological Lagrangians\cite{Kaymakcalan:1983qq, Oh:2000qr, Casalbuoni:1996pg, Colangelo:2002mj}
\begin{eqnarray}
&&\mathcal{L}_{D^{\ast }D\pi}=-\frac{ig_{D^\ast DP}}{\sqrt{2}}{D^{\dagger}\partial_{\mu}\vec\pi\cdot\vec\tau D^{\ast \mu}},\nonumber\\
&&\mathcal{L}_{D^\ast D\eta}=-ig_{D^\ast D\eta}D^{\dagger}\partial^{\mu}\eta D^{\ast }_{\mu},\nonumber\\
&&\mathcal{L}_{D_s^\ast DK}=ig_{D^\ast DP}(D_s^{\ast -\mu}D\partial_{\mu} K^{\dagger}),\nonumber\\
&&\mathcal{L}_{D^{\ast }D^\ast \pi}=\frac{1}{2\sqrt{2}}g_{D^\ast D^\ast P}\epsilon_{\mu\nu\alpha\beta}{D^{\ast \dagger\mu}
   \partial^{\nu}\vec\pi\cdot\vec\tau \lrpartial{}^\alpha D^{\ast \beta}},\nonumber\\
&&\mathcal{L}_{D^\ast D^\ast \eta}=g_{D^\ast D^\ast \eta}
    \epsilon_{\mu\nu\alpha\beta}{D^{\ast \dagger\mu}\partial^{\nu}
    \eta\lrpartial{}^{\alpha}D^{\ast \beta}},\nonumber\\
&&\mathcal{L}_{D_s^\ast D^\ast K}=\frac{1}{2}g_{D^\ast D^\ast P}
    \epsilon_{\mu\nu\alpha\beta}{D_s^{\ast -\mu}\partial^{\nu}K^{\dagger}
    \lrpartial{}\partial^{\alpha}D^{\ast \beta}},\nonumber\\
&&\mathcal{L}_{D_s^\ast D^\ast K^\ast }=ig_{D^\ast D^\ast V}D_s^{\ast -\nu}\lrpartial{}^{\mu}D_{\nu}^{\ast }
    K^{\ast \dagger}_{\mu}\nonumber\\
&&\phantom{\mathcal{L}_{D_s^\ast D^\ast K^\ast }=}+4if_{D^\ast D^\ast V}D^{\ast -}_{s\mu}(\partial^{\mu}K^{\ast \dagger\nu}-\partial^{\nu}K^{\ast \dagger\mu})D^{\ast }_{\nu},\nonumber\\
&&\mathcal{L}_{K^\ast K\pi}=ig_{K^\ast K\pi} K_{\mu}^{\ast \dagger}\vec\pi\cdot\vec\tau\lrpartial{}^{\mu}K,\\
&&\mathcal{L}_{K^\ast K\eta}=-ig_{K^\ast K\eta}K_{\mu}^{\ast \dagger}\eta\lrpartial{}^{\mu}K,
\label{Eq:Lag1}
\end{eqnarray}
where $A\lrpartial{}B\equiv A (\partial B)-(\partial A) B$, $\vec\tau$ is the Pauli matrix, $\vec\pi$ represents the pion triplets, and $K^{(\ast)}$ and $D^{(\ast)}$ are the doublets of strange and charmed mesons, respectively,
\begin{equation}
K^{(\ast )}=\left(\begin{array}{c}
K^{(\ast )+}\\
K^{(\ast )0}
\end{array}\right)\,,\qquad
D^{(\ast )}=\left(\begin{array}{c}
D^{(\ast )0}\\
D^{(\ast )+}
\end{array}\right)\,.
\end{equation}

In the heavy quark-limit, the coupling constants $g_{D^\ast D^{(\ast)} P}$ could be related to the gauge coupling constant $g$ via
\begin{eqnarray}
g_{D^\ast D^\ast P}=\frac{2g}{f_{\pi}}\,,\quad g_{D^\ast DP}=\frac{2g}{f_{\pi}}\sqrt{m_{D^\ast }m_D}\,,
\end{eqnarray}
where $f_{\pi}=132$\,MeV is the decay constant of the pion and the gauge coupling $g=0.59$ is estimated from the experimental value of the partial width for the $D^{\ast+} \to D^+ \pi^0$. The involved coupling constants of $K^\ast$ are \cite{Oh:2000qr},
\begin{eqnarray}
g_{D_s^\ast D^\ast K^\ast}=\frac{\beta g_V}{\sqrt{2}}, \ \ \ f_{D_s^\ast D^\ast K^\ast} =\frac{\lambda g_V}{\sqrt{2}} \sqrt{m_{D_s^\ast D^\ast}},
\end{eqnarray}
where the gauge couplings $\beta=0.9$, $\lambda=0.56$ and $g_V=m_\rho/f_\pi$. As for the coupling constants of $g_{K^\ast K \pi}$ and $g_{K^\ast K \eta}$, we adopt $g_{K^\ast K \pi}=3.21$ and $g_{K^\ast K \eta}=4.47$, which are evaluated by SU(3) symmetry \cite{Liu:2005jb}.

The involved interaction related to the photon field and the charmed mesons is in the form \cite{Chen:2010re},
\begin{eqnarray}
&&\mathcal{L}_{D^\ast D\gamma}=\bigg\{\frac{g_{D^{\ast +}D^+\gamma}}{4}e\epsilon^{\mu\nu\alpha\beta}F_{\mu\nu}D^{\ast +}_{\alpha\beta}D^-\nonumber\\
&&\quad \phantom{\mathcal{L}_{D^\ast D\gamma}=}+\frac{g_{D^{\ast 0}D^0\gamma}}{4}e\epsilon^{\mu\nu\alpha\beta}F_{\mu\nu}D^{\ast 0}_{\alpha\beta}\bar D^0\bigg\}+H.c.\,,
\end{eqnarray}
where the field-strength tensors are defined as $F_{\mu\nu}= \partial_{\mu}A_{\nu} -\partial_{\nu}A_{\mu},\  D^\ast _{\alpha \beta} =\partial_{\alpha}D^\ast _{\beta} -\partial_{\beta}D^\ast _{\alpha}$. The coupling constant $g_{D^{\ast+} D^+ \gamma }={}-0.5 \ \mathrm{GeV}^{-1}$ is estimated from the partial width of $D^{\ast +} \to D^+ \gamma$: the minus sign is adopted according to the lattice QCD and QCD sum rule calculations  \cite{Becirevic:2009xp,Zhu:1996qy}. As for $g_{D^{\ast0} D^0 \gamma}$, only the branching ratios of the $ D^{\ast0} \to D^0 \gamma  $ and $D^{\ast0} \to D^0 \pi^0 $ are measured. Here, we can roughly estimate the partial width of the $D^{\ast0} \to D^0 \pi^0$ from that of the $D^{\ast +} \to D^+ \pi^0$ via isospin symmetry \cite{Dong:2008gb, Chen:2015igx}. With the measured ratio of the $\Gamma(D^{\ast0} \to D^0 \gamma)$ and $\Gamma(D^{\ast0} \to D^0 \pi^0)$, we can obtain the partial width of the $D^{\ast0}\to D^0 \gamma$ and the corresponding coupling constant $g_{D^{\ast 0} D^0 \gamma}$ as  $g_{D^{\ast 0} D^0 \gamma} =2.0\ \mathrm{GeV}^{-1}$.

In the present work, the decays of the $D_{s1}(2460) \to D_{s0}^\ast(2317) \pi^0$ and $D_{s1}(2460) \to D_{s} \pi^0$ are the isospin-violating processes, which are also contributed from the $\eta-\pi^0$ mixing. The $\eta-\pi^0$ mixing scheme is in the form \cite{Gasser:1984gg},
\begin{equation}
\mathcal{L}_{\eta\pi^0}=\mu\frac{(m_d-m_u)}{\sqrt{3}}\pi^0\eta,
\end{equation}
where $m_u$ and $m_d$ are the current quark masses of the $u$ and $d$ quarks, respectively, and $\mu$ is the condensate parameter.

\subsection{The decay of $D_{s1}(2460)\to D_{s0}^\ast(2317) \pi^0$}

\begin{figure}[htb]
\begin{tabular}{cc}
 \includegraphics[scale=0.38]{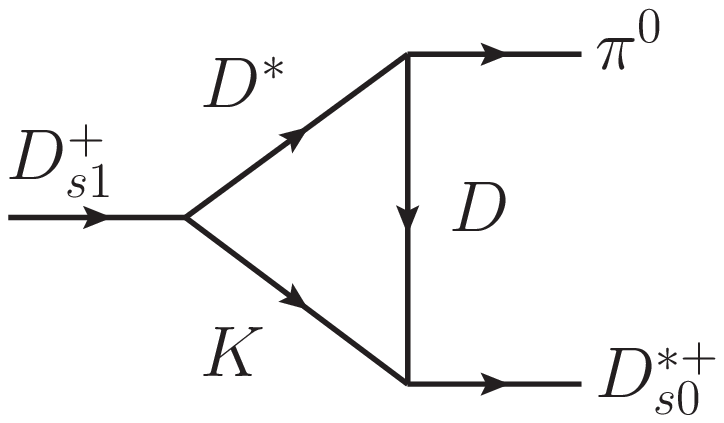}
&\includegraphics[scale=0.38]{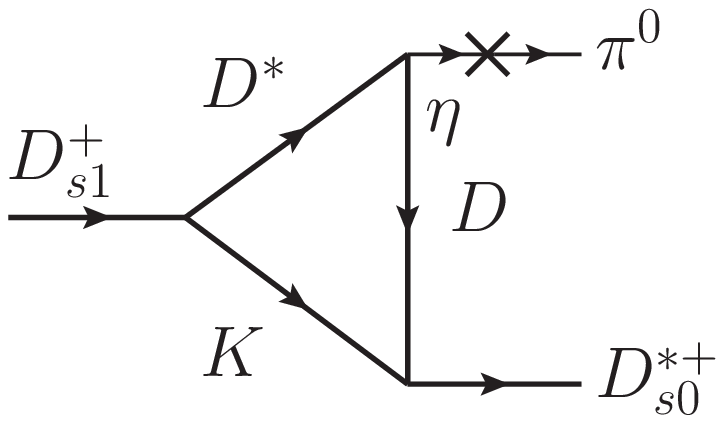}\\
(a) &(b)
\end{tabular}
\caption{Diagrams contributing to the pionic transition from the $D_{s1}(2460)$ to the $D_{s0}^\ast(2317)$. Diagram (a) is the direct contribution and diagram (b) is the contribution from $\eta-\pi^0$ mixing. \label{fig:feyn-pi-a}}
\end{figure}

The decay of the $D_{s1}(2460)\to D_{s0}^\ast(2317) \pi^0$ occurs via a subprocess $D^\ast \to D \pi^0$ in the hadronic molecular picture, and the hadronic-level description of this process is presented in Fig. \ref{fig:feyn-pi-a}(a). Since this decay is an isospin-violating process, we also include the contribution from the $\eta-\pi^0$ mixing as presented in Fig. \ref{fig:feyn-pi-a}(b). With the effective interactions listed above, we can get the amplitude corresponding to Fig. \ref{fig:feyn-pi-a}(a) as
\begin{eqnarray}
\mathcal{M}_{a}&=& (i)^3 \int \frac{d^4q}{(2\pi)^4} \big[g_{D_{s1}D^\ast K}\epsilon^{\phi}_{D_{s1}}\tilde\Phi_{D_{s1}}(-P_{12}^2,\Lambda^2_{D_{s1}})\Big] \nonumber\\
&&\times \Big[g_{D^\ast _{s0}DK}\tilde\Phi_{D_{s0}^\ast }(-P_{20}^2,\Lambda^2_{D_{s0}^\ast })\Big] \ \Big[\frac{ig_{D^\ast DP}}{\sqrt{2}}(-ip_3^{\mu})\Big]  \nonumber\\
&&\times  \frac{-g^{\phi\mu}+p_1^{\phi}p_1^{\mu}/m_1^2}{p_1^2-m_1^2}
    \frac{1}{p_2^2-m_2^2}
    \frac{1}{q^2-m_q^2},\label{Eq:pi-a}
\end{eqnarray}
where $P_{12}=(p_1w_{D^\ast K}-p_2w_{KD^\ast })$ and $P_{20}=qw_{DK}-p_2w_{KD}$. The amplitude related to Fig. \ref{fig:feyn-pi-a}(b) is,
\begin{eqnarray}
\mathcal{M}_b&=&(i)^3 \int \frac{d^4q}{(2\pi)^4} \big[g_{D_{s1}D^\ast K}\epsilon^{\phi}_{D_{s1}}\tilde\Phi_{D_{s1}}(-P_{12}^2,\Lambda^2_{D_{s1}})\Big] \nonumber\\
&&\times \Big[g_{D^\ast _{s0}DK}\tilde\Phi_{D_{s0}^\ast }(-P_{20}^2,\Lambda^2_{D_{s0}^\ast })\Big] \ \Big[\frac{ig_{D^\ast D\eta}}{\sqrt{2}}(-ip_3^{\mu})\Big]  \nonumber\\
&&\times  \frac{-g^{\phi\mu}+p_1^{\phi}p_1^{\mu}/m_1^2}{p_1^2-m_1^2}
    \frac{1}{p_2^2-m_2^2}
    \frac{1}{q^2-m_q^2}\nonumber\\
&&\times\mu\frac{m_d-m_u}{\sqrt{3}}\frac{1}{m_{\pi}^2-m_{\eta}^2},\label{Eq:pi-b}
\end{eqnarray}
where $m_{\pi}^2=(m_u+m_d)\mu$, $m_{\eta}^2 = \frac{2}{3} (m+2m_s)\mu$ and $m=(m_u+m_d)/2$. The above amplitude $\mathcal{M}_b$ can be reduced to
\begin{eqnarray}
&&\mathcal{M}_b=\mathcal{M}_a|_{\pi^0 \to \eta}\frac{\sqrt{3}}{4}\frac{(m_d-m_u)}{(m_s-m)},
\end{eqnarray}
where $\mathcal{M}_a|_{\pi^0 \to \eta}$ indicates the amplitude obtained by replacing the related coupling constants of $\pi^0$ with those of $\eta$. The total amplitude of the $D_{s1}(2460) \to D_{s0}^\ast(2317) \pi^0$ is
\begin{eqnarray}
\mathcal{M}_{D_{s1}\to D_{s0}^\ast \pi^0} =\mathcal{M}_a +\mathcal{M}_b. \label{eq:amp1}
\end{eqnarray}

\subsection{The decay of $D_{s1}(2460) \to D_{s0}^\ast(2317) \gamma$}

\begin{figure}[htb]
\begin{tabular}{cc}
\includegraphics[scale=0.38]{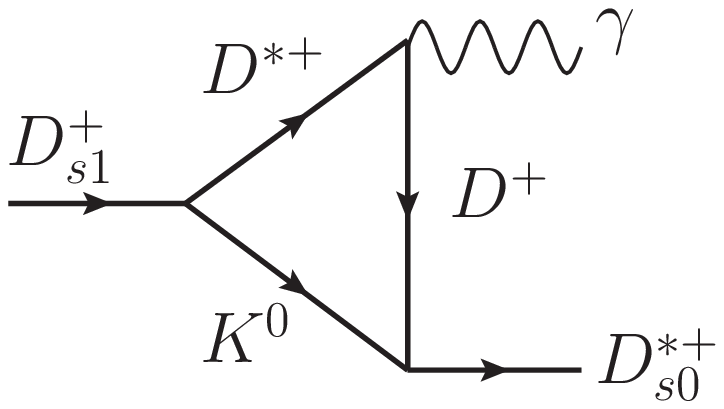}
&\includegraphics[scale=0.38]{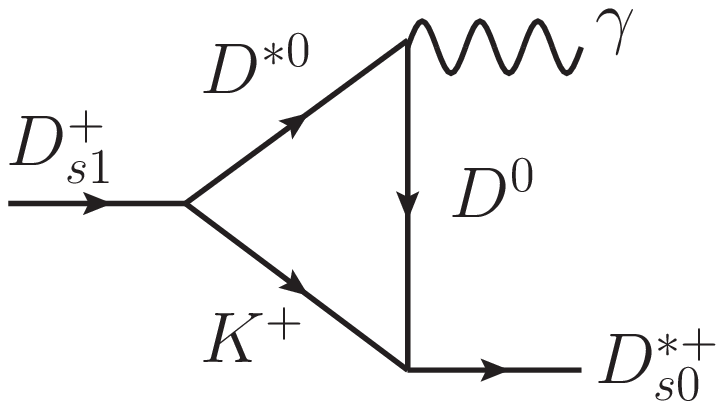}\\
(a) &(b)
\end{tabular}
\caption{Diagrams contributing to the radiative transition from the $D_{s1}(2460)$ to the $D_{s0}^\ast(2317)$. (a) is the contribution from charged charmed mesons and (b) is the contribution from the
neutral charmed mesons.. \label{fig:feyn-gamma}}
\end{figure}

As for the decay of $D_{s1}(2460) \to D_{s0}^\ast(2317) \gamma$, it occurs via the subprocess $D^\ast \to D \gamma $ as shown in Fig. \ref{fig:feyn-gamma}. With the effective Lagrangians given above, we can obtain the amplitude corresponding to Fig. \ref{fig:feyn-gamma}(a) as
\begin{eqnarray}
\mathcal{M}_{a}  &=&
   (i)^3 \int \frac{d^4q}{(2\pi)^4}\Big[g_{D_{s1}D^\ast K}\epsilon^{\phi}_{D_{s1}}\tilde\Phi_{D_{s1}}(-P_{12}^2,\Lambda^2_{D_{s1}})\Big] \nonumber\\
   && \times  \Big[g_{D^\ast _{s0}DK}\tilde\Phi_{D_{s0}^\ast }(-P_{20}^2,\Lambda^2_{D_{s0}^\ast })\Big]\Big[\frac{e g_{D^{\ast +}D^+\gamma}}{4} \nonumber\\
&& \times
\epsilon_{\mu\nu\alpha\beta}\epsilon^{\eta}_{\gamma}(ip_3^{\nu}g^{\mu\eta}-ip_3^{\nu}g^{\nu\eta})
    (ip_1^{\alpha}g^{\beta\tau}-ip_1^{\beta}g^{\alpha\tau})\Big]\nonumber\\
&&\times
    \frac{-g^{\phi\tau}+p_1^{\phi}p_1^{\tau}/m_1^2}{p_1^2-m_1^2}
    \frac{1}{p_2^2-m_2^2}
    \frac{1}{q^2-m_q^2}.
\end{eqnarray}
As for the amplitude corresponding to Fig. \ref{fig:feyn-gamma}(b), it can be obtained from  the above amplitude by replacing the masses and coupling constants with those in Fig. \ref{fig:feyn-gamma}(b), i.e.,
\begin{eqnarray}
\mathcal{M}_{b} = \mathcal{M}_{a}\left|_{g_{D^{\ast +} D^+ \gamma} \to g_{D^{\ast 0} D^0 \gamma} }^{m_{D^{\ast +}} \to m_{D^{\ast 0}}, m_{D^+} \to m_{D^0}, m_{K^0} \to m_{K^+}}\right..
\end{eqnarray}
Then the total amplitude for $D_{s1}(2460) \to D_{s0}^\ast(2317) \gamma$ is
\begin{eqnarray}
\mathcal{M}_{D_{s1} \to D_{s0}^\ast \gamma} =\mathcal{M}_{a}+\mathcal{M}_b. \label{eq:amp2}
\end{eqnarray}
It should be noticed that after performing the loop integral, the above amplitude can be reduced to the form,
\begin{eqnarray}
 \mathcal{M}_{D_{s1} \to D_{s0}^\ast \gamma}= g_{D_{s1} D_{s0}^\ast \gamma} \varepsilon_{\mu \nu \alpha \beta} \epsilon_{D_{s1}}^\mu \epsilon_\gamma^\nu p_\gamma^\alpha p_{D_{s1}}^\beta,
\end{eqnarray}
which is obviously gauge invariant and the coupling constant $g_{D_{s1} D_{s0}^\ast \gamma}$ could be estimated from the amplitude in Eq. (\ref{eq:amp2}).

\subsection{The decay of $D_{s1}(2460) \to D_s^\ast \pi^0$}

\begin{figure}[h]
\begin{tabular}{ccc}
\includegraphics[scale=0.35]{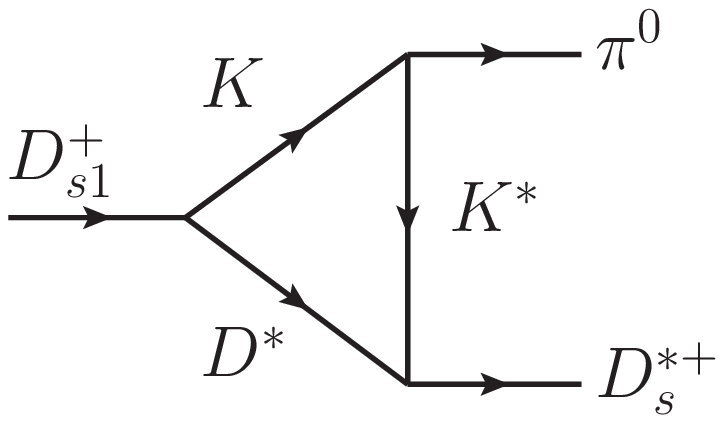}
&\includegraphics[scale=0.35]{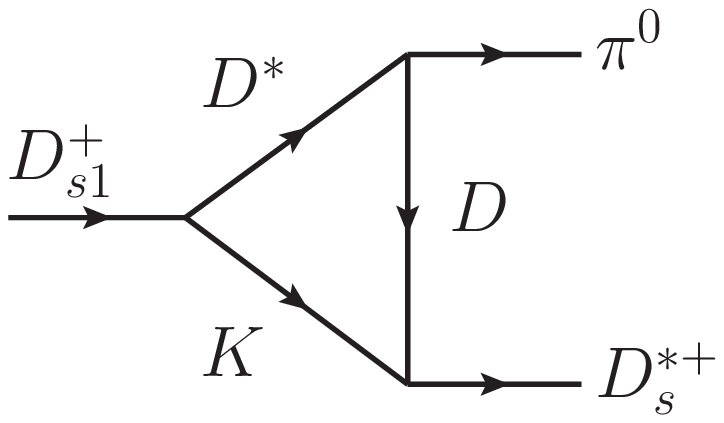}
&\includegraphics[scale=0.35]{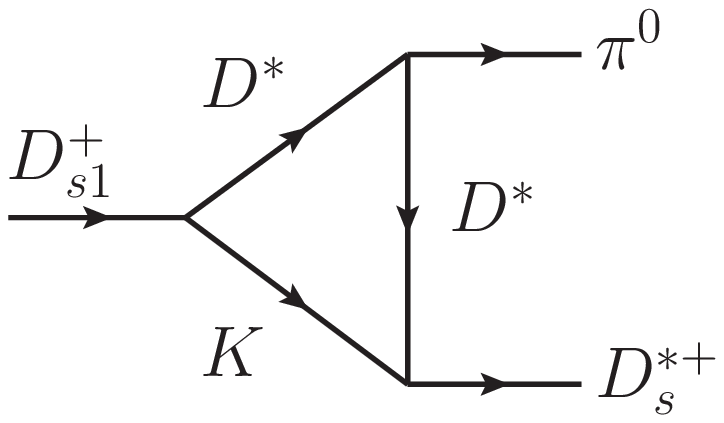}\\
(a) & (b) & (c)\\
\includegraphics[scale=0.35]{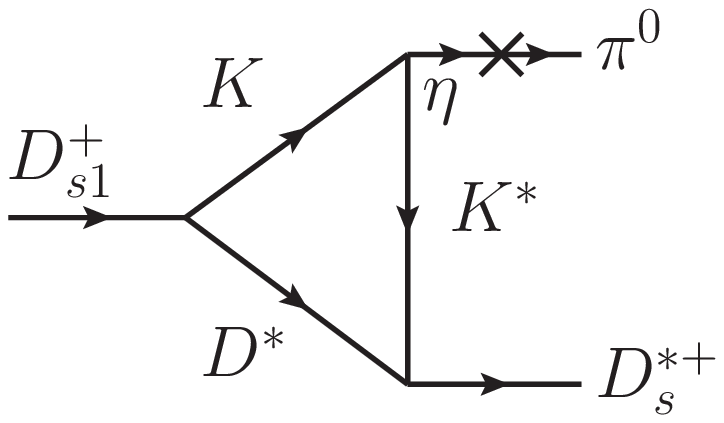}
&\includegraphics[scale=0.35]{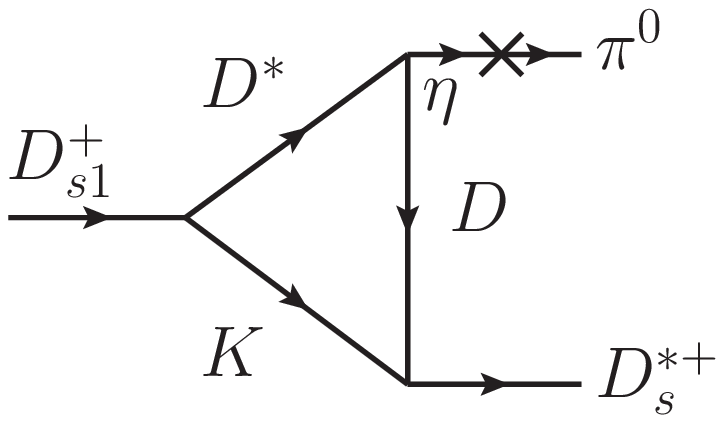}
&\includegraphics[scale=0.35]{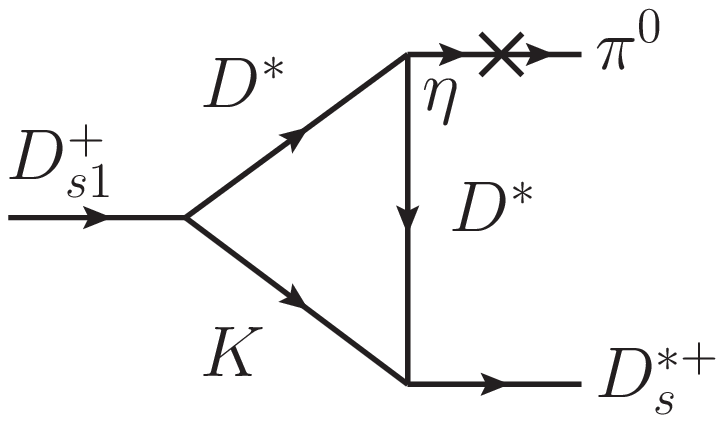}\\
(d) & (e) & (f)
\end{tabular}
\caption{Diagrams contributing to process $D_{s1}^+\to D_s^{\ast +}\,\pi^0$. Diagrams (a), (b) and (c) are direct processes, where the $\pi^0$ directly couples to strange mesons or charmed mesons. Diagrams (d), (e) and (f) are indirect processes, where $\pi^0$ couples to  strange mesons or charmed mesons via $\eta-\pi^0$ mixing.\label{fig:feyn-sstar} }
\end{figure}

We can estimate the partial width of $D_{s1}(2460) \to D_{s}^\ast \pi^0$ and compare the evaluated ratio of the $\Gamma(D_{s1}(2460) \to D_{s0}^\ast \gamma)$ and $\Gamma(D_{s1}(2460) \to D_{s}^\ast \pi^0)$ to further test the hadronic molecular interpretations of the $D_{s1}(2460)$ and $D_{s0}^\ast(2317)$. Similar to the process $D_{s1}(2460) \to D_{s0}^\ast \pi^0$, the decay of $D_{s1}(2460) \to D_s^\ast \pi^0$ is also an isospin-violating process, which also arises from the direct $\pi^0$ coupling and $\eta-\pi^0$ mixing as shown in Fig.~\ref{fig:feyn-sstar}. In our calculations, in addition to the diagrams considered in Ref.~\cite{Faessler:2007us}, we include the diagrams due to the $D^\ast D^\ast \pi$ and $D^\ast D^\ast \eta$ interactions. The concrete forms of the amplitudes corresponding to Figs. \ref{fig:feyn-sstar}(a)--\ref{fig:feyn-sstar}(c) are
\begin{eqnarray}
{\mathcal{M}_a} &=&
    (i)^3\int \frac{d^4q}{(2\pi)^4}\Big[g_{D_{s1}D^\ast K}\epsilon^{\phi}_{D_{s1}}\tilde \Phi_{D_{s1}}(-P_{12}^2,\Lambda^2_{D_{s1}})\Big] \nonumber\\
  &&\times \Big[ig_{K^\ast K\pi}(ip_1^{\eta}+ip_3^{\eta})\Big] \Big[ig_{D^\ast D^\ast V}\epsilon^{\tau}_{D_s^\ast }g^{\tau\rho}
    (ip_2^{\sigma}+ip_4^{\sigma})\nonumber\\
&&+4if_{D^\ast D^\ast V}\epsilon^{\tau}_{D_s^\ast }(iq^{\tau}g^{\rho\sigma}-iq^{\rho}g^{\tau\sigma})\Big]
    \frac{1}{p_1^2-m_1^2}\nonumber\\
&& \times
    \frac{-g^{\rho\phi}+p_2^{\rho}p_2^{\phi}/m_2^2}{p_2^2-m_2^2}
    \frac{-g^{\eta\sigma}+q^{\eta}q^{\sigma}/m_q^2}{q^2-m_q^2}\,,
\end{eqnarray}
\begin{eqnarray}
{\mathcal{M}_b} &=&
    (i)^3\int \frac{d^4q}{(2\pi)^4}\Big[g_{D_{s1}D^\ast K}\epsilon^{\phi}_{D_{s1}} \tilde\Phi_{D_{s1}}(-P_{12}^2,\Lambda^2_{D_{s1}})\Big] \nonumber\\
&&\times  \Big[\frac{-ig_{D^\ast DP}}{\sqrt{2}}(-ip_3^{\mu})\Big]
    \Big[ig_{D^\ast DP}\epsilon^{\nu}_{D_s^\ast }(ip_2^{\nu})\Big]\nonumber\\
&&\times\frac{-g^{\phi\mu}+p_1^{\mu}p_1^{\phi}/m_1^2}{p_1^2-m_1^2}
    \frac{1}{p_2^2-m_2^2}
    \frac{1}{q^2-m_q^2}\,,\\
%
%
{\mathcal{M}_c} &=&
    (i)^3\int \frac{d^4q}{(2\pi)^4}\Big[g_{D_{s1}D^\ast K}\epsilon^{\phi}_{D_{s1}} \tilde\Phi_{D_{s1}}(-P_{12}^2,\Lambda^2_{D_{s1}})\Big]\nonumber\\
  &&  \times\Big[\frac{1}{2\sqrt{2}}g_{D^\ast D^\ast P}\varepsilon_{\eta\tau\rho\sigma}(-ip_3^{\tau})(ip_1^{\rho}+iq^{\rho})\Big]\nonumber\\
&&\times
    \Big[\frac{1}{2}g_{D^\ast D^\ast P}\varepsilon_{\mu\nu\alpha\beta}\epsilon_{D_s^\ast }^{\mu}(ip_2^{\nu})(iq^{\alpha}+ip_4^{\alpha})\Big]
    \nonumber\\&&\times\frac{-g^{\sigma\phi}+p_1^{\sigma}p_1^{\phi}/m_1^2}{p_1^2-m_1^2}
     \frac{1}{p_2^2-m_2^2}
    \frac{-g^{\eta\beta}+q^{\eta}q^{\beta}/m_q^2}{q^2-m_q^2}\,.
\end{eqnarray}
As for the contributions from $\eta-\pi^0$ mixing, the amplitudes corresponding to Figs. \ref{fig:feyn-sstar}(d)-\ref{fig:feyn-sstar}(f) by,
\begin{eqnarray}
\mathcal{M}_d&=&\mathcal{M}_a|_{\pi^0 \to \eta} \frac{\sqrt{3}}{4}\frac{(m_d-m_u)}{(m_s-m)},\nonumber\\
\mathcal{M}_e&=&\mathcal{M}_b|_{\pi^0 \to \eta} \frac{\sqrt{3}}{4}\frac{(m_d-m_u)}{(m_s-m)},\nonumber\\
\mathcal{M}_f&=&\mathcal{M}_c|_{\pi^0 \to \eta} \frac{\sqrt{3}}{4}\frac{(m_d-m_u)}{(m_s-m)}.
\end{eqnarray}
The total amplitude of $D_{s1}(2460) \to D_{s}^\ast \pi^0$ is
\begin{eqnarray}
\mathcal{M}_{D_{s1}\to D_s^\ast \gamma}= \sum_{n=a}^{f} \mathcal{M}_n.
\label{eq:amp3}
\end{eqnarray}

With the total amplitudes defined in Eqs. (\ref{eq:amp1}), (\ref{eq:amp2}) and (\ref{eq:amp3}), one can estimate the partial width by,
\begin{eqnarray}
\Gamma=\frac{1}{3}\frac{1}{8\pi}\frac{|\vec p|}{m_{D_{s1}}^2}|\overline{\mathcal{M}}|^2,
\end{eqnarray}
where $\vec p$ is the momentum of the final state in the $D_{s1}(2460)$ rest frame and the overline indicates sum over polarizations of vector mesons.

\section{Numerical results} \label{sec4}
\begin{table}[hbt]
\caption{The masses of the involved particles in units of GeV \cite{Agashe:2014kda}.\label{tab:mass}}
\begin{tabular}{cccccccc}
\toprule[1pt]
State & Mass & State & Mass & State & Mass & State & Mass \\
\midrule[1pt]
$D^0$ &1.8648  & $D^\pm$    &1.8696  &$D^{\ast0}$   &2.0069  &$D^{\ast \pm}$  &2.0102\\
$K^0$ &0.4976  & $K^{\pm}$  &0.4936 &$K^{\ast0}$   &0.8958  &$K^{\ast \pm}$  &0.8916 \\
$D_s^{\ast \pm}$   &2.1121  &$D_{s0}^{\ast \pm}$   &2.3177  &$D_{s1}^{\pm}$  & 2.4595   &$\pi^0$ &0.1349 \\
$\eta$  &0.5478 \\
\bottomrule[1pt]
\end{tabular}
\end{table}
\begin{figure}[htb]
\includegraphics[scale=0.75]{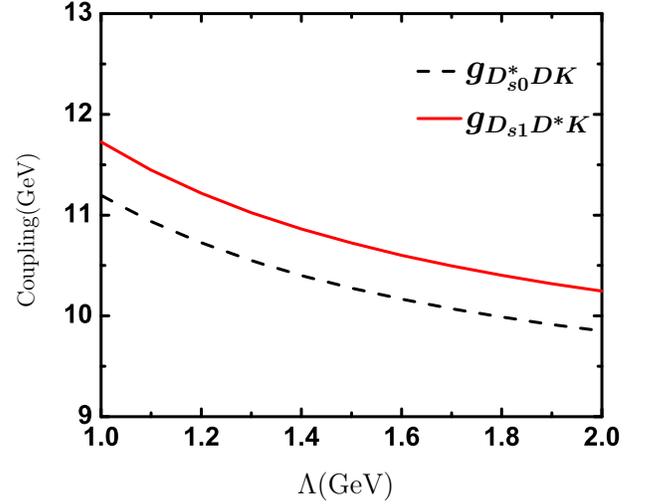}
\caption{ The $\Lambda$ dependence of the coupling constants $g_{D_{s0}^\ast DK}$  and $g_{D_{s1}D^\ast K}$, where $\Lambda_{D_{s1}}=\Lambda_{D_{s0}^\ast}=\Lambda$.\label{fig:n-gds0}}
\end{figure}

All the masses of the involved particles are listed in Table~\ref{tab:mass}.
Besides the coupling constants discussed in Sec.~\ref{Sec3}, the coupling constants of $D_{s1}(2460)/D_{s0}^\ast(2317)$ to their components could be estimated by the compositeness conditions given by Eq.~\eqref{eq:cp1}. The phenomenological parameters $\Lambda_{D_{s1}}$ and $\Lambda_{D_{s0}^\ast}$ are of order 1 GeV. Here, we vary the parameters $\Lambda_{D_{s1}}$ and  $\Lambda_{D_{s0}^\ast}$ from 1 to 2 GeV \cite{Faessler:2007gv, Faessler:2007us}. The $\Lambda_{D_{s1}} = \Lambda_{D_{s0}^\ast} = \Lambda$ dependences of the coupling constants $g_{D_{s1} D^\ast K}$ and $g_{D_{s0}^\ast DK}$ are presented in Fig.~\ref{fig:n-gds0}. These two coupling constants monotonously decrease with the increasing of the parameter $\Lambda$. In particular, the coupling constants $g_{D_{s1} D^\ast K}$ and $g_{D_{s0}^\ast DK}$ decrease from  $11.73$ to  $10.25$ GeV and from $11.20$ to $9.85$ GeV, respectively, when $\Lambda $ increases from 1 to 2 GeV.

\begin{figure}[htb]
\includegraphics[scale=0.75]{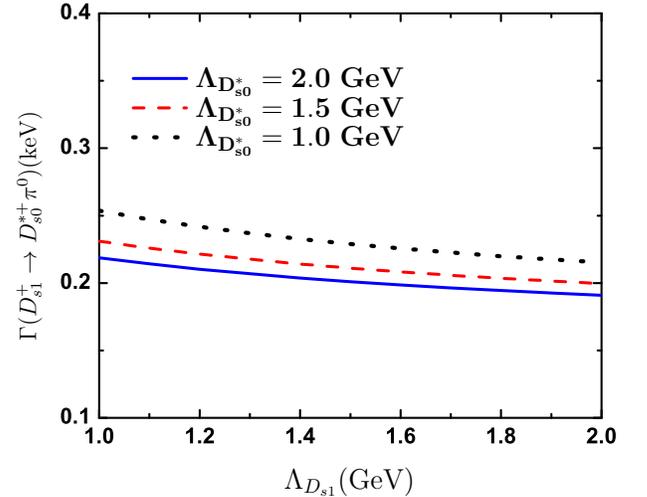}
\caption{. The $\Lambda_{D_{s1}}$ dependence of the decay width for $D_{s1}(2460) \to D^{\ast }_{s0}(2317) \pi^0$. \label{fig:n-s0-pi}}.
\end{figure}

The partial width of the $D_{s1}(2460) \to D_{s0}^\ast (2317)\pi^0$ is presented in Fig.~\ref{fig:n-s0-pi}. In the present calculation, we vary the $\Lambda_{D_{s1}}$ from 1.0 to 2.0 GeV and take typical values of $\Lambda_{D_{s0}^\ast} = 1.0, 1.5$ and $2.0$\,GeV. Our calculations indicate that the partial width of the $D_{s1}(2460) \to D_{s0}^\ast (2317)\pi^0$ is of order 0.1 keV, which is rather small since the phase space of this process is very limited. In addition, this partial width weakly depends on the parameters $\Lambda_{D_{s1}}$ and $\Lambda_{D_{s0}^\ast}$, and decreases with the increasing of $\Lambda_{D_{s1}}$ or $\Lambda_{D_{s0}^\ast}$. In the case of $\Lambda_{D_{s0}^\ast}=1.0$ GeV, the partial width for the $D_{s1}(2460) \to D_{s0}^\ast (2317)\pi^0$ decreases from $0.25$ to $0.21$ keV with $\Lambda_{D_{s1}}$ increasing from 1.0 to 2.0 GeV. In the considered parameter region, the partial width for the $D_{s1}(2460) \to D_{s0}^\ast (2317)\pi^0$ is predicted to be about $0.19 \sim 0.25$ keV.

\begin{figure}[htb]
\includegraphics[scale=0.75]{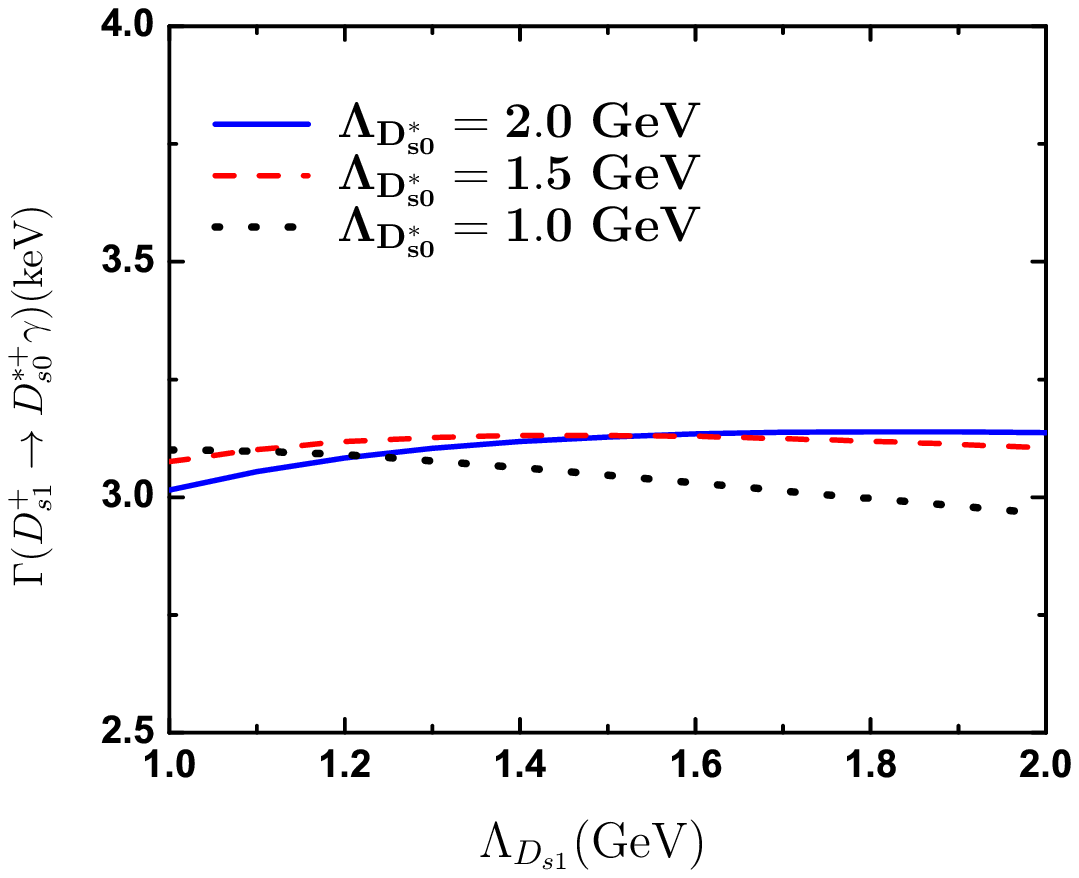}
\caption{  The same as Fig. \ref{fig:n-s0-pi} but for $D_{s1}(2460)\to D^{\ast }_{s0}(2317)\gamma$ process. \label{fig:n-s0-ga}}.
\end{figure}

The $\Lambda_{D_{s1}}$ dependence of the partial width for the $D_{s1}(2460) \to D_{s0}^\ast(2317) \gamma$ is presented in Fig. \ref{fig:n-s0-ga}. Similar to the pionic transition from the $D_{s1}(2460)$ to the $D_{s0}^\ast(2317)$, the partial width for the $D_{s1}(2460) \to D_{s0}^\ast(2317) \gamma$ also weakly depends on the parameters $\Lambda_{D_{s1}}$ and $\Lambda_{D_{s0}^\ast}$. In the considered parameter region, the partial width for the $D_{s1}(2460) \to D_{s0}^\ast(2317) \gamma$ varies from  $2.96$ to $3.13$ keV. The PDG average of the branching ratio of the $D_{s1}(2460) \to D_{s0}^\ast(2317) \gamma$ is $3.7_{-2.4}^{+5.0} \%$. However, the width of $D_{s1}(2460)$ is not well determined, as one cannot compare the theoretical value of the partial width with the experimental measurement. Here, we also notice that both widths for the $D_{s1}(2460) \to D_{s0}^\ast(2317) \pi^0$ and $D_{s1}(2460) \to D_{s0}^\ast(2317) \gamma$ weakly depend on the model parameters, and the former one is about 1 order smaller than the latter one, which indicates that the branching ratio of $D_{s1}(2460) \to D_{s0}^\ast(2317) \pi^0$ should be of order $10^{-3}$.

\begin{figure}[htb]
\includegraphics[scale=0.75]{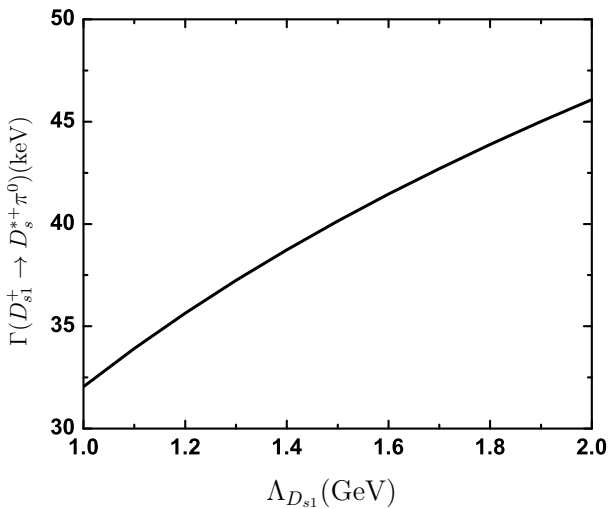}
\caption{The $\Lambda_{D_{s1}}$ dependence of the partial width for $D_{s1}(2460)\to D^{\ast }_s\pi^0$. \label{fig:n-s1-sstar}}.
\end{figure}

In Fig. \ref{fig:n-s1-sstar}, we present the $\Lambda_{D_{s1}}$ dependence of the partial width for the $D_{s1}(2460) \to D_{s}^{\ast} \pi^0$ which increases with the increasing of $\Lambda_{D_{s1}}$. In particular, the partial width varies from 32 to 46 keV with $\Lambda_{D_{s1}}$ increasing from 1.0 to 2.0 GeV, which is much larger than the partial widths for the $D_{s0}^\ast(2317) \gamma$ and $D_{s0}^\ast(2317) \pi^0$ modes. In addition, the partial widths for the $D_{s1}(2460) \to D_{s0}^\ast(2317) \gamma$ and $D_{s1}(2460) \to D_{s}^\ast \pi^0$ have been estimated in the present work, and the ratio of $\Gamma(D_{s1}(2460) \to D_{s0}^\ast(2317) \gamma)$ and $\Gamma(D_{s1}(2460) \to D_{s}^\ast \pi^0)$ is estimated to be $(6.6-10.2)\times 10^{-2}$ in the considered parameter region, which is safely under the upper limit reported by the CLEO and {\it BABAR} collaborations \cite{Besson:2003cp, Aubert:2003pe}.

\begin{table}[htb]
\caption{A comparison of the the partial widths (in units of keV) from different models. \label{Tab:Com}}
\begin{tabular}{cccccc}
\toprule[1pt]
Channel & Present & Ref. \cite{Bardeen:2003kt} & Ref. \cite{Colangelo:2005hv} & Ref. \cite{Godfrey:2003kg} & Ref. \cite{Song:2015nia}\\
\midrule[1pt]
$D_{s1} \to D^\ast_{s0} \gamma$ & $3.0 \sim 3.1$  & 2.74 & $0.5 \sim 0.8 $ & $0.012$ & $\cdots$ \\
$D_{s1} \to D^\ast_{s0} \pi^0$  & $0.19 \sim 0.22$ & 0.0079 & $\cdots$ & $\cdots$ & $\cdots$ \\
$D_{s1} \to D_s^\ast \pi^0$     & $31.3 \sim 45.2$ & 21.5  & $\cdots$ & $\sim 10$  & $11.9$ \\
\bottomrule[1pt]
\end{tabular}
\end{table}

In Table \ref{Tab:Com}, we collect our estimates of the partial widths of the $D_{s1}(2460) \to D_{s0}^\ast(2317) \gamma,\ D^\ast_{s0}(2317) \pi^0$, and $D_s^\ast \pi^0$ and compare with the results evaluated in the $P$-wave charmed-strange meson scheme. In Ref. \cite{Bardeen:2003kt}, the decays of the $D_{s1}(2460)$ were estimated in a full chiral theory and the partial widths for the $D_{s1}(2460) \to D_{s0}^\ast \gamma$ and $D_{s1}(2460) \to D^\ast \pi^0$ are very similar to the present results obtained in a molecular scenario, but for the $D_{s1}(2460) \to D_{s0}^\ast(2317) \pi^0$ mode, the results from Ref. \cite{Bardeen:2003kt} are much smaller than the present one. The light-cone sum rule calculation for $D_{s1}(2460)\to D_{s0}^\ast(2317) \gamma$ is about 20\% of that obtained in the present calculation \cite{Colangelo:2005hv}. The estimations in the relativistic quark model indicated that the partial widths of $D_{s1}(2460) \to D_{s0}^\ast \gamma$ and $D_{s1}(2460) \to D_{s}^\ast \pi^0$ were 0.012 and about 10 keV, respectively \cite{Godfrey:2003kg, Song:2015nia}, which are rather different with the results in the present work.

\section{Summary} \label{sec5}

In the present work, we estimated the partial widths for the radiative and pionic transitions from the $D_{s1}(2460)$ to the $D_{s0}^\ast(2317)$ in a molecular scenario, in which the $D_{s1}(2460)$ and the $D_{s0}^\ast(2317)$ are assigned as a $DK$ and a $D^\ast K$ hadronic molecule, respectively. To further test the molecular interpretations of the $D_{s1}(2460)$ and the $D_{s0}^\ast(2317)$, we also calculated the partial width for $D_{s1}(2460) \to D_{s}^\ast \pi^0$. In the considered parameter region, the partial widths are evaluated to be
\begin{eqnarray}
\Gamma(D_{s1}(2460)\to D^{\ast }_{s0}(2317)\pi^0)&=& 0.19\sim 0.22\,\text{keV},\nonumber\\
\Gamma(D_{s1}(2460)\to D^{\ast }_{s0}(2317)\gamma)&=&3.0 \sim 3.1\,\text{keV},\nonumber\\
\Gamma(D_{s1}(2460)\to D^{\ast }_s\,\pi^0) &=&31.3 \sim 45.2\,\text{keV}.
\end{eqnarray}

Our estimates indicate that the partial width for the $D_{s1}(2460)\to D^{\ast }_{s0}(2317)\pi^0$ is about 1 order smaller than that of $D_{s1}(2460)\to D^{\ast }_{s0}(2317)\gamma$. The branching ratio for $D_{s1}(2460)\to D^{\ast }_{s0}(2317)\gamma$ is measured to be $3.7 _{-2.4}^{+5.0}\%$ \cite{Agashe:2014kda}, and thus the branching ratio for $D_{s1}(2460)\to D^{\ast }_{s0}(2317)\pi^0$ is roughly determined to be of order $10^{-3}$.
In addition, we further estimate the ratio of ${\Gamma(D_{s1}(2460) \to D_{s0}^{\ast }(2317)\gamma)}$ and ${\Gamma(D_{s1}(2460) \to D_s^{\ast +}\,\pi^0)}$ to be
\begin{eqnarray}
\frac{\Gamma(D_{s1}(2460) \to D_{s0}^{\ast }(2317)\gamma)}{\Gamma(D_{s1}(2460) \to D_s^{\ast +}\,\pi^0)}=(6.6-10.6)\times 10^{-2},
\end{eqnarray}
which is consistent with the experimental measurements from the CLEO and {\it BABAR} collaborations \cite{Besson:2003cp, Aubert:2003pe}.

At present, the experimental information on the $D_{s1}(2460)$ and $D_{s0}(2317)$ is still not abundant. In particular, the widths of these states are not well determined. The measurements of their decay behaviors at LHCb and the forthcoming Belle II could provide a further test to the results in the present work.

\section* {Acknowledgements}
The work of D.-Y. C. is supported by the National Natural Science Foundation of China under Grant No. 11375240. Y.-L. M. is supported in part by the National Science Foundation of China (NSFC) under Grant No. 11475071, 11547308 and the Seeds
Funding of Jilin University.

\end{document}